\documentclass[reprint,amsmath,amssymb,pra,aps,noeprint,floatfix, nolinenumbers,superscriptaddress,longbibliography]{revtex4-2}

\usepackage{graphicx}% Include figure files
\usepackage{dcolumn}% Align table columns on decimal point
\usepackage{bm}% bold math
\usepackage{hyperref}% add hypertext capabilities
\usepackage{physics}
\usepackage{xcolor}
\hypersetup{
    unicode=true, % non-Latin characters in Acrobat?s bookmarks
    plainpages=false,
    colorlinks=true,% false: boxed links; true: colored links
    linkcolor=blue,% color of internal links
    citecolor=blue,% color of links to bibliography
    filecolor=blue,% color of file links
    urlcolor=blue% color of external links
}

\usepackage{txfonts} % Nicer fonts?
\newcommand{\spinvec}{|\langle \hat{\vb{F}} \rangle|}
\newcommand{\freeze}[1]{\tilde{#1}}

\begin{document}

\title{Dynamics of a Nonequilibrium  Discontinuous Quantum Phase Transition\\ in a Spinor Bose-Einstein Condensate}

\author{Matthew T. Wheeler}
\affiliation{School of Engineering, Mathematics and Physics, University of East Anglia, Norwich Research Park, Norwich, NR4 7TJ, United Kingdom}
\affiliation{Centre for Photonics and Quantum Science, University of East Anglia, Norwich Research Park, Norwich, NR4 7TJ, United Kingdom}
\author{Hayder Salman}
\affiliation{School of Engineering, Mathematics and Physics, University of East Anglia, Norwich Research Park, Norwich, NR4 7TJ, United Kingdom}
\affiliation{Centre for Photonics and Quantum Science, University of East Anglia, Norwich Research Park, Norwich, NR4 7TJ, United Kingdom}
\author{Magnus O. Borgh}
\affiliation{School of Engineering, Mathematics and Physics, University of East Anglia, Norwich Research Park, Norwich, NR4 7TJ, United Kingdom}
\affiliation{Centre for Photonics and Quantum Science, University of East Anglia, Norwich Research Park, Norwich, NR4 7TJ, United Kingdom}

\begin{abstract}%
Symmetry-breaking quantum phase transitions lead to the production of topological defects or domain walls in a wide range of physical systems. In second-order transitions, these exhibit universal scaling laws described by the Kibble-Zurek mechanism, but for first-order transitions a similarly universal approach is still lacking.
Here we propose a spinor Bose-Einstein condensate as a testbed system where critical scaling behavior in a first-order quantum phase transition can be understood from generic properties.
We demonstrate the applicability of the Kibble-Zurek mechanism for this transition to determine the critical exponents for: (1) the onset of the decay of the metastable state on short times scales, and (2) the number of resulting phase-separated ferromagnetic domains at longer times, as a one-dimensional spin-1 condensate is ramped across a first-order quantum phase transition. The predictions are in excellent agreement with mean-field numerical simulations and provide a paradigm for studying the decay of metastable states in experimentally accessible systems.
\end{abstract}

\maketitle

\section{Introduction}
Classical and quantum nonequilibrium phase transitions arise in many areas of physics, ranging from cosmology~\cite{Kibble1980,Mazumdar2019}, to condensed matter~\cite{Chuang1991,Hendry1994,Bauerle1996,Ruutu1996,Sondhi1997,Polkovnikov2011}, and to ultracold atomic gases~\cite{Hadzibabic2006,Langen2015,Fletcher2015,Liu2018}.
For a second-order (continuous) phase transition, a correlation length and time scale can be identified that characterise the coherence and dynamical response of the system.  As the critical point is approached,  both of these exhibit power-law divergences described by critical exponents~\cite{Hohenberg1977,Goldenfeld1992}. In non-equilibrium phase transitions, this implies that close to the critical point, the system is no longer able to adiabatically follow the ground state~\cite{Dziarmaga2010,DelCampo2014}.  Causally disconnected regions then choose the new broken symmetry state independently, which results in the formation of topological defects or domain boundaries at a density related to the quench rate of the control parameter.  

The Kibble-Zurek mechanism (KZM) provides a theoretical framework that can predict the density of these defects or domain walls for a finite quench rate from universal properties of the continuous phase transition. First introduced by Kibble in the context of cosmology as a mechanism for the formation of cosmic strings in the early universe~\cite{Kibble1976, Kibble1980}, it was subsequently extended by Zurek to condensed matter systems~\cite{Zurek1985, Zurek1993, Zurek1996}. It has since been successfully verified in many settings, including thermally driven transitions~\cite{Su2013,Lamporesi2013,Liu2014,Donadello2016, Beugnon2017} and quantum phase transitions (QPTs)~\cite{Dziarmaga2005, Damski2005,Yi2020}, and has been demonstrated to apply to quantum-annealing implementations of quantum computation~\cite{King2022,King2023}.

Recently, there has been increasing interest in studying first-order QPTs~\cite{Coulamy2017,Shimizu2018,Pelissetto2018,Pelissetto2020,Sinha2021}, where metastability plays a crucial role, including in cold-atom systems~\cite{Billam2022,Song2022,Zenesini2023}. A classical example of such metastability is the transition of supercooled water which remains liquid below the freezing point. For first-order quantum phase transitions, such `supercooling'-like behaviour can lead to a zero-temperature false vacuum. This state plays an important role in particle physics and cosmology~\cite{Coleman1977,Fialko2015,Devoto2022}, but an understanding of how the metastable state decays is hampered by the lack of a general theoretical approach dealing with first-order QPTs. The KZM has been partially explored for discontinuous transitions in, e.g., the transverse-field Ising model~\cite{Coulamy2017,Pelissetto2018,Sinha2021} and the Bose-Hubbard model~\cite{Shimizu2018}. The recently predicted modification of scaling behavior as first-order characteristics are introduced to a non-equilibrium classical phase transition~\cite{Suzuki2024} further raises the question of when the KZM can be expected to apply in discontinuous QPTs.

Here we propose an experimentally accessible nonequilibrium first-order QPT where the decay of metastable states can be understood through the KZM. In particular, we study the persistence of the metastable state following a finite quench of the quadratic Zeeman shift in a spin-1 Bose-Einstein condensate (BEC) with ferromagnetic (FM) interactions. We demonstrate that the onset of decay of the metastable state representing the false vacuum agrees with the critical scaling law predicted by a generalisation of the KZM to our first-order QPT.

A key feature of this phase transition in a ferromagnetic spinor BEC is the formation of phase-separated domains at long times past the transition point. Therefore, in addition to the short-time behaviour characterising the decay of the metastable state, we also apply the KZM to determine the scaling of the number of phase-separated domains at later times. We show that the KZM accurately predicts the scaling of the number of domains for fast to intermediate quench rates, whereas for slow quenches, deviations appear similarly to some second-order QPTs~\cite{Damski2005,Zurek2005}.

\section{Results and Discussion}

\subsection{Mean-field theory of the spin-1 BEC}

Focusing on an ultracold atomic system has the advantage that
the QPT is easier to control for isolated systems.
Atomic BECs in particular are pristine systems and offer a highly controllable platform where the strength of inter-atomic interactions and the confining trapping potentials can be tuned.  Consequently, they are already popular example systems for phase-transition experiments~\cite{Hadzibabic2006,Kruger2007,Weiler2008,Chomaz2015}, as well as nonequilibrium physics, even in low dimensions, ranging from relaxation dynamics~\cite{Gring2012,Reeves2022} to quantum quenches~\cite{Sadler2006,Barnett2011,Navon2015,Symes2017,Kang2017,Prufer2018,Schmied2019,Liu2020, Kirkby2024}.  
Unlike scalar BECs,  the spin degrees of freedom are not frozen out in spinor BECs.  
These additional degrees of freedom give rise to a non-trivial phase diagram even at zero temperature~\cite{Kawaguchi2012,Stamper-Kurn2013,Murata2007,Borgh2014,Matuszewski2009,Mirkhalaf2021} and a correspondingly rich array of topological defects and textures~\cite{Kawaguchi2012,Stamper-Kurn2013,Leanhardt2003,Sadler2006,Seo2015,Kang2019,Weiss2019,Xiao2022}. 
For these reasons, studying non-equilibrium dynamics and QPTs with spinor BECs has attracted much attention~\cite{Sadler2006,Damski2006,Damski2007, Lamacraft2007,Saito2007, Saito2007a,Vengalattore2008,Swislocki2013, Witkowska2013, Anquez2016,Williamson2016,Kang2017,Prufer2024,Huh2024}.  

As our example system, we consider a spin-1 BEC 
described by the mean-field condensate spinor wave function $\Psi = (\psi_1, \psi_0, \psi_{-1})^T$.
The Hamiltonian density then reads~\cite{Kawaguchi2012}
\begin{equation}
    H = H_0 + \frac{c_0}{2}n^2 + \frac{c_1}{2}n^2\spinvec^2 - pn\langle \hat{F_z} \rangle + qn\langle\hat{F}_z^2\rangle,
    \label{eq: Hamiltonian-density}
\end{equation}
where $H_0=(\hbar^2/2M)|\nabla \Psi|^2 + nV(z)$ for atomic mass $M$ and external trapping potential $V(z)$.
Here, $n=\sum_m\psi_m^*\psi_m$ is the total atomic density.
The condensate spin operator $\hat{\vb{F}}\equiv (\hat{F}_x, \hat{F}_y, \hat{F}_z)$ is the vector of spin-1 Pauli-type matrices such that $\langle\hat{F}_\mu\rangle=\tfrac{1}{n}\sum_{mm'}\psi_m^*(\hat{F}_\mu)_{mm'}\psi_{m'}$ for $\mu = x,y,z$. Conservation of angular momentum in $s$-wave scattering means that the longitudinal magnetisation $M_z = \int \langle\hat{F}_z\rangle \, dz$ is conserved on experimental time scales.
The spin-independent and -dependent interaction strengths arise from the $s$-wave scattering lengths $a_\mathcal{F}$ in the spin-$\mathcal{F}$ channels of colliding spin-1 atoms as $c_0=4\pi\hbar^2(a_0+2a_2)/3M$ and $c_1=4\pi\hbar^2(a_2-a_0)/3M$, respectively.
Linear and quadratic Zeeman shifts of strengths $p$ and $q$, respectively, may arise from an applied magnetic field along the $z$-direction, or in the latter case be induced by an AC Stark shift~\cite{Gerbier2006,Santos2007}, which gives precise control over both strength and sign.  
Due to conservation of $M_z$, a uniform linear Zeeman shift only causes precession of the spin, under which the Hamiltonian is invariant. We therefore only consider effects from the quadratic Zeeman shift.

We consider atoms with $c_1<0$, such as $^{87}$Rb~\cite{Klausen2001} or $^7$Li~\cite{Huh2020,Huh2024}, which provide an interesting phase diagram arising from the competition between the third and last term in Eq.~\eqref{eq: Hamiltonian-density}.
A three-component broken-axisymmetry (BA) phase with zero longitudinal magnetisation occurs for $0 < Q = q / (|c_1|n_0) < 2$~\cite{Kawaguchi2012} where $n_0$ is the background density
in a uniform system:
\begin{equation}
    \psi_{\pm 1} = \frac{\sqrt{2n_0}}{4}\sqrt{2-Q}, \qquad \psi_0 = \frac{\sqrt{n_0}}{2}\sqrt{2 + Q}.
    \label{eq: BA-GS}
\end{equation}
In addition, a ferromagnetic (FM) state occurs for $Q<0$: $\Psi = (\sqrt{n_0}, 0, 0)^T$ or $\Psi = (0, 0, \sqrt{n_0})^T$.
Since $M_z$ is conserved, a BA initial condition with $M_z=0$ implies that the FM phase results in the formation of phase-separated domains with opposite spin projection as $Q$ is ramped across the phase transition. 
The associated instability that leads to the emergence of phase separated domains when $c_1<0$ is not captured in the single-mode approximation~\cite{Matuszewski2009,Mirkhalaf2021}.

\subsection{Theory of Kibble-Zurek scaling}

In contrast to previous studies on QPTs, the first-order QPT between the BA and FM phases in a spin-1 BEC with FM interactions 
corresponds to a discontinuous quantum critical point (DQCP)~\cite{Suzuki2015}, the quantum analogue of the classical discontinuous critical point~\cite{Nauenberg1975A,Fisher1982}. As it does not meet the general criteria of applicability of the standard KZ theory, the KZM has hitherto been little studied in this context.
We consider, in particular, a one-dimensional (1D) spin-1 BEC with FM interactions in a ring-trap geometry.
By quenching the quadratic Zeeman shift, the system can transition from the BA phase into a phase-separated FM phase where domains of atoms with opposite condensate-spin projection form~\cite{Kawaguchi2012}.
Unlike the single-mode scenario in an antiferromagnetic condensate~\cite{Qiu2020}, where there is no domain formation, here phase separation is a consequence of the FM interactions under conservation of longitudinal magnetisation in a spatially extended BEC. Its 1D nature, however, has the advantage that once the domains form, they are frozen in and cannot undergo any coarsening dynamics~\cite{Sabbatini2011,Sabbatini2012}, which facilitates the accurate analysis of the scaling behaviour predicted by the KZM for the DQCP.

Moreover, the transition point between the BA and FM phases at $(q_c,p_c)=(0,0)$ is a DQCP. It
satisfies five conditions~\cite{Suzuki2015}, which permit scaling arguments to be applied. A critical point at $q=q_c$ and $p=p_c$, where $p$ functions as a symmetry-breaking field, is a DQCP of this kind if
(1) the energy density $\epsilon(q,p)$ across the transition is continuous, $\epsilon(q_c^+, p_c) - \epsilon(q_c^- , p_c) = 0$, but (2) its derivative is discontinuous, $\partial\epsilon(q_c^+, p_c)/\partial q - \partial\epsilon(q_c^-, p_c)/\partial q \ne 0$ (establishing its first-order nature). Here, $q_c^+$ and $q_c^-$ correspond to approaching $q_c$ from above or below,  respectively. Further, (3) the order parameter $m=-\partial \epsilon(q, p)/\partial p$ must exhibit a discontinuous jump with respect to $q$ as the critical point is crossed, $|m(q_c^- , p_c)| > |m(q_c^+, p_c)| = 0$, and additionally, (4) the order parameter is also discontinuous with respect to $p$: $|m(q_c,  p_c^{\pm})| > 0$, ensuring that the DQCP is not a triple point in the $(p,q)$ parameter space~\cite{Fisher1982}. Finally,  we require (5) that the derivative of the energy be bounded as the critical point is approached: $|\partial \epsilon(q_c^{\pm},  p_c)/\partial q| < \infty$. Criteria (1)--(5) permit us to investigate the KZM for our DQCP.

The energy densities per particle for the BA and FM states are, respectively~\cite{Kawaguchi2012}, 
\begin{align}
    \epsilon_{\mathrm{BA}} &= \frac{(-p^2+q^2+2qc_1 n_0)^2}{8c_1 n_0 q^2} + \frac{1}{2}c_0 n_0,\\
    \epsilon_{\mathrm{FM}_{\pm}} &= \mp p + q + \frac{1}{2} n_0(c_0 + c_1),
\end{align} 
where the subscript $+(-)$ denotes the FM phase with spin pointing up (down).
These energies are continuous at the critical point $(q_c,p_c)=(0,0)$.
The derivatives, however, are discontinuous, but remain bounded.
Meanwhile, the relevant order parameter for the BA and FM states is $m_\mathrm{BA}=p(p^2-q^2-2qc_1n_0)/8c_1 n_0 q$ and $m_{\mathrm{FM}_{\pm}}=\pm 1$, respectively, which is precisely the local magnetisation $F_z=|\psi_1|^2-|\psi_{-1}|^2$ in both phases~\cite{Kawaguchi2012}.
For $p_c=0$ this order parameter is zero in the BA phase and becomes non-zero in the FM phase.
A non-zero $p$, however, causes the order parameter to be non-zero in both phases.
This means that the BA to FM transition satisfies all the conditions for a DQCP.

We now recall the key arguments of the KZM as applied to QPTs~\cite{Dziarmaga2010}.  A continuous second-order phase transition can be characterised by the divergence of a single instantaneous correlation length $\xi \sim |q(t)-q_c|^{-\nu}$ and a single instantaneous relaxation time $\tau \sim [\xi(t)]^z$, where $\nu$ and $z$ are the correlation-length and dynamical critical exponents, respectively. 
In the standard QPT scenario~\cite{Dziarmaga2010,Damski2005}, the relaxation time is usually set by the inverse of the energy gap $\Delta$ between the ground state and the first excited state of a gapped mode~\cite{Zurek2005, Damski2006}: $\tau \simeq \Delta^{-1}$.
A system initially prepared in the ground state follows this state adiabatically as long as the relaxation time remains small.
However, as the critical point is approached,  the  energy gap vanishes and the relaxation time diverges, which leads to the breakdown of the adiabatic regime.
This divides the dynamics into three stages: adiabatic, frozen, and adiabatic again as the system crosses the critical point.
Assuming a quench of the form $|q(t)-q_c| \sim |t/\tau_Q|$,  where $\tau_Q^{-1}$ is the quench rate,
the freezing time is estimated to be $|\, \freeze{t}\,| = \tau(\freeze{t}\,)$. This leads to a scaling for the freezing time given by
\begin{equation}
    |\,\freeze{t}\,| \sim \tau_Q^{z\nu/(z\nu + 1)}.
    \label{eq: KZ-freezing-scaling}
\end{equation}
It follows from Eq.~\eqref{eq: KZ-freezing-scaling} and the scaling of the correlation length that at the freezing time,  $\freeze{\xi} \sim \xi(\,\freeze{t}\,) \sim \tau_Q^{\nu/(z\nu + 1)}$. This provides an estimate for the number of defects or domains: $N_{\text{dom}} \sim \freeze{\xi}^{-d} \sim \tau_Q^{-d\nu/(z\nu + 1)}$.

Note that once scaling behaviour is assumed, Eq.~\eqref{eq: KZ-freezing-scaling} can be derived by assuming the existence of single length and time scales, which can typically be justified for second-order transitions. It does not always hold for discontinuous phase transitions but can be extended for a DQCP satisfying the five criteria. These key aspects of the KZM are therefore generic and also do not depend on the existence of a gapped mode (indeed, Zurek considered a thermal superfluid phase transition characterised by a gapless dispersion~\cite{Zurek1985}). However, the specific values of the critical exponents are determined by the form of the dispersion relation near the critical point.

\begin{figure}
    \centering
    \includegraphics[width=\columnwidth]{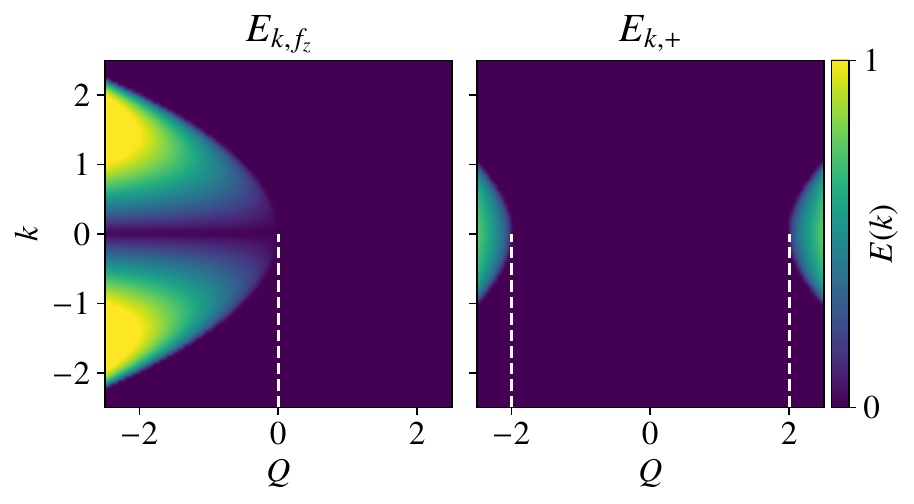}
    \caption{\label{fig: bogoliubov-energies}\textbf{Stability of Bogoliubov modes.} Imaginary parts of \(E_{k, f_z}\)
    and \(E_{k, +}\) are shown as functions of the wavenumber $k$ and quadratic level shift $Q$ (colour scale). \(E_{k, f_z}\) is unstable (positive imaginary part) for $Q<0$, corresponding to the broken axisymmetry to ferromagnetic discontinuous quantum critical point, while \(E_{k, +}\) instead becomes unstable at the broken axisymmetry to polar second-order transition at \(Q = 2\).
    }
\end{figure}

For a first-order transition, one must consider the Bogoliubov modes corresponding to the phase from which the transition is approached~\cite{Suzuki2015}, here the BA phase. The three spin-1 Bogoliubov modes~\cite{Uchino2010} then correspond to spin waves $E_{\bf{k},f_z}$, density waves $E_{\bf{k},+}$, and the so-called theta modes $E_{\bf{k},-}$, (see ``Appendix'').
Only $E_{\vb{k}, {+}}$ is gapped in the long-wavelength limit and determines the KZM scaling for the second-order transition from the polar to the BA phase~\cite{Damski2007} (where modes on either side of the transition may be used).
By contrast, the relevant mode for the BA to FM transition is
\begin{equation}
    E_{k, f_z}(k) = \sqrt{\epsilon_{k}(\epsilon_{k} + q)},
    \label{eq: bogoliubov-energy}
\end{equation}
where $\epsilon_{k} = \hbar^2k^2/2M$ is the kinetic energy. This spectrum is gapless in the long-wavelength limit. The imaginary part of $E_{k, f_z}$ together with that of $E_{\vb{k}, {+}}$ is shown in Fig.~\ref{fig: bogoliubov-energies} and clearly illustrates that an instability can occur at $Q=0$ only for modes with $k \ne 0$. These unstable modes are responsible for the formation of phase-separated domains in the FM phase. In contrast, the most unstable mode for the phase transition at $Q=2$ corresponds to $k=0$ within the range $1<Q<2$.

To derive KZ scaling for a QPT with a gapless mode~\cite{Suzuki2015} we consider the more general spectrum, $E_k^2 \sim |q(t)-q_c|^{\alpha} \epsilon_k^{\eta}+\epsilon_k^{2z}$, of which Eq.~\eqref{eq: bogoliubov-energy} is a special case.  
The correlation length associated with the transition can then be inferred by equating the two terms. Therefore, to find scaling solutions consistent with the KZM where $E_k \sim |q(t)-q_c|^{z}$, we assume that the two terms are of equal magnitude and scale similarly
by making the ansatz $k \sim |q(t)-q_c|^{\nu}$ (corresponding to $k\sim\xi^{-1}$) to derive the condition $\alpha = \nu(2z - \eta)$.
Subsequently, the adiabatic-impulse approximation states that the impulse regime begins when $E_k^2 = \dot{E}_k$. This yields
\begin{equation}
    |\,\freeze{t}\,| \sim \tau_Q^{\alpha/(2+\alpha)}\freeze{k}^{-\eta/(2+\alpha)} \sim \tau_Q^{\nu z/(1+\nu z)} 
    \label{eq:ad-imp}
\end{equation}
for the freezing time upon using the above scaling assumed for $k$.
This immediately implies the characteristic momentum scale
    $\freeze{k} \sim \tau_Q^{-\nu/(z\nu + 1)}$
from which we extract the defect density
    $N_{\text{dom}} \sim \freeze{k}^d \sim \tau_Q^{-d\nu/(z\nu + 1)}$,
where $d$ is the dimensionality of the system.

We have thus recovered the same general expression for the KZ scaling as for a continuous phase transition. This should not be surprising since the order of the transition does not enter into the derivations of Eqs.~\eqref{eq: KZ-freezing-scaling} and \eqref{eq:ad-imp} once the single length and time scales are established. In addition to these universally valid expressions, deriving the scaling law from the form of the dispersion relation also allows us to determine the critical exponents.

Specifically for our 1D system ($d=1$ and $q_c = 0$), Eq.~\eqref{eq: bogoliubov-energy} implies $\alpha=1$, $\eta=2$ and $z=2$. This is equivalent to setting $z=2$ and $\nu=1/2$ 
corresponding to a defect-density scaling
\begin{equation}
    N_{\text{dom}} \sim \tau_Q^{-1/4}.
    \label{eq: n_d-scaling}
\end{equation}
Therefore, despite originating in the same model Hamiltonian,  this scaling is clearly different from that found for the KZM in continuous phase transitions through a QCP in spinor BECs~\cite{Damski2007,Saito2007,Saito2007a,Saito2013,Swislocki2013}. Our results thus indicate a new scaling regime, compared with the polar-to-BA QPT in the same system. The difference can be attributed to the fact that the most unstable mode in the latter case corresponds to $k=0$, while in our case it acquires a $k$-dependence.

\subsection{Numerical results}

To check our prediction,  we numerically evolve the time-dependent spin-1 Gross-Pitaevskii equations (GPEs) obtained from Eq.~\eqref{eq: Hamiltonian-density}
using a symplectic algorithm~\cite{Symes2016} (see ``Appendix'') and interaction strengths $c_0/c_1 = -20$. This choice reduces the disparity in system timescales and hence the computational overheads. All results have also been verified for $^{87}$Rb ($c_0/c_1 = -216$)  $^7$Li ($c_0/c_1 = -2.17$) parameters to confirm that the KZM should be observable in existing experiments.
\begin{figure}
    \centering
    \includegraphics[width=\columnwidth]{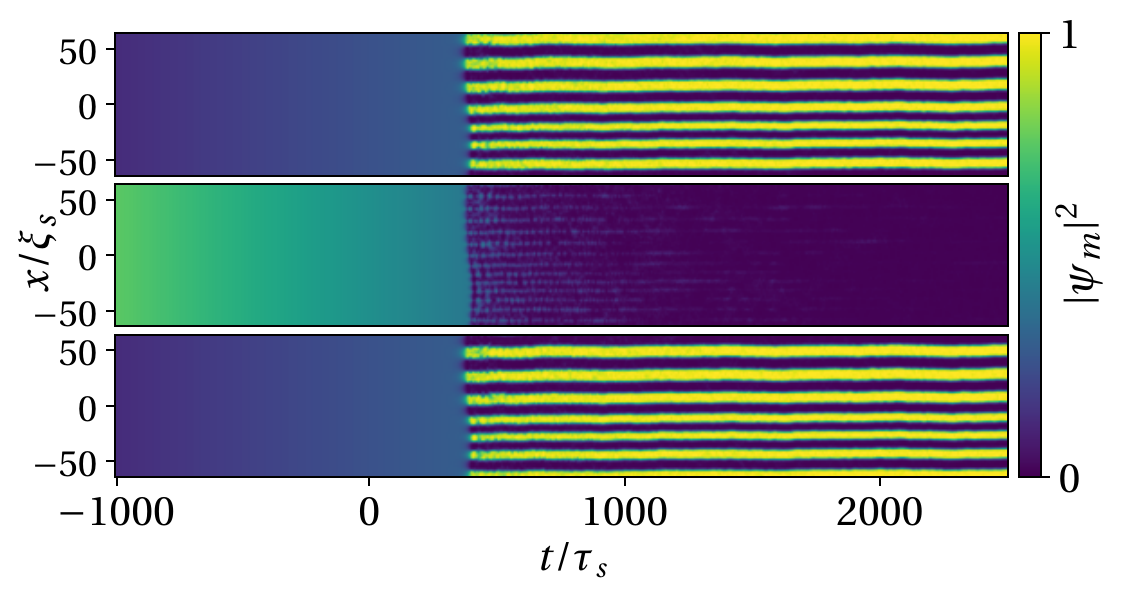}
    \caption{\textbf{Formation of ferromagnetic domains.} Component densities (colour scale) in a $128\xi_s$ subregion of the $\psi_1$ (top), $\psi_0$ (middle) and $\psi_{-1}$ (bottom) components for a typical simulation with the quench time $\tau_Q = 1000$.}
    \label{fig: component-densities}
\end{figure}
Typical results for $\tau_Q=1000$ are shown in Fig.~\ref{fig: component-densities}.
We see the clear formation of FM domains after crossing the critical point at $t=0$.
\begin{figure}
    \centering
    \includegraphics[width=\columnwidth]{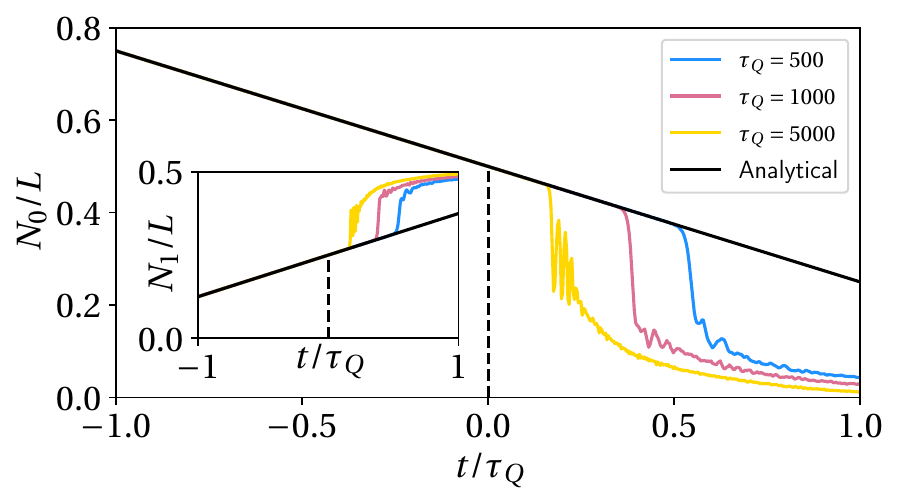}
    \caption{\textbf{Decay of the metastable state.} Normalised atom number $N_0/L$ for the $\psi_0$ component using the analytical prediction (black line) in Eq.~\eqref{eq: BA-GS}. Also plotted are numerically calculated values for quench times $\tau_Q=500, 1000, 5000$ (blue, red and yellow lines, respectively).
    Inset: Normalised atom number for the $\psi_1$ component using the same analytical and numerical data as the main figure. 
    We note that $N_1/L$ approaches 0.5 since the longitudinal magnetisation $M_z=0$ is conserved. The critical point is marked with a black dashed line.}
    \label{fig: density-deviation}
\end{figure}
Figure \ref{fig: density-deviation} shows the normalised atom number for the $\psi_0$ component as the system is quenched for various $\tau_Q$.
Initially, the system tracks the BA-phase ground state~\cite{Qiu2020}, Eq.~\eqref{eq: BA-GS}.
After passing the critical point, the system is no longer able to adiabatically track the true ground state. Rather, it evolves in  a metastable BA state, even for $t/\tau_Q>0$, until it emerges from the impulse regime at a time clearly dependent on the quench rate. At this point, the metastable state decays with an associated abrupt drop in the density of the $\psi_0$ component, signalling a discontinuous phase transition to the FM phase. 
This coincides with an increase in the $\psi_{\pm 1}$ components, where the FM domains start to form (Fig.~\ref{fig: density-deviation} inset).

\begin{figure}
    \centering
    \includegraphics[width=\columnwidth]{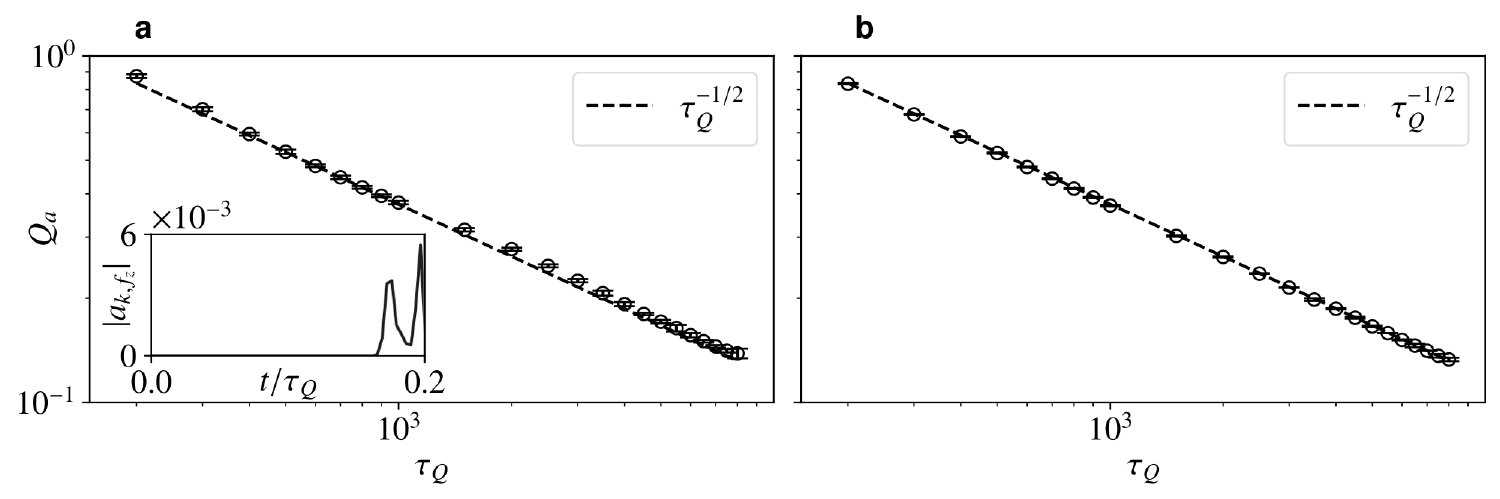}
    \caption{\textbf{Kibble-Zurek scaling of the freezing time.} \textbf{a.} Scaling of the critical value $Q_a = |Q(\tilde{t}\,)|$, where $\tilde{t}$ is the freezing time, versus the quench time $\tau_Q$.
    Overlaid is the power-law scaling $\tau_Q^{-1/2}$ (dashed line).
    Inset: the quantity $|a_{k, {f_z}}|$ for $\tau_Q=5000$ and the $n=1$ mode (i.e.\ $k=2\pi/L)$. 
    \textbf{b.} Same but with \(Q_a\) calculated using the deviation from the
    analytically predicted value of \(\psi_0\) (see ``Appendix'') reproducing the same scaling.
    For both methods, each point is averaged over 50 runs, and the error bars represent
    $\pm 1$ standard deviation.
    }
    \label{fig:Q_a-scaling}
\end{figure}

To determine the freezing time as well as the short time scaling behaviour, we introduce $\hat{a}_{k, \pm 1}$, defined as the Fourier transforms of the $\psi_{\pm 1}$ components, respectively.
Since the transition to the FM phase results in phase-separated domains, driven by an instability associated with a Bogoliubov mode related to $\hat{a}_{k, {f_z}} = (\hat{a}_{k, 1} - \hat{a}_{k, -1})/\sqrt{2}$ (see ``Appendix''), we extract the critical time $\freeze{t}$ such that $|\hat{a}_{k, {f_z}}(\freeze{t}\,)| = 0.01 \times \mathrm{max}\{|\hat{a}_{k, {f_z}}(t)| : t\}$.
The critical Zeeman value  is defined as $Q_a = |Q(\freeze{t}\,)|$.
The inset in Fig.~\ref{fig:Q_a-scaling} shows the typical growth of $|\hat{a}_{k, {f_z}}|$, which demonstrates that it remains zero until the system passes the critical point. Thereafter, it undergoes growth with strong oscillations.
The choice of $0.01$ when extracting $\freeze{t}$ is arbitrary, but we find qualitatively similar results in tests with values up to $0.1$.

Figure~\ref{fig:Q_a-scaling} reveals a clear $Q_a\sim\tau_Q^{-1/2}$ power law for a large range of $\tau_Q$.  
Despite the discontinuous nature of the phase transition,  scaling behaviour is still observed~\cite{Turban2002,Continentino2004, Continentino2017}. However, this scaling is different from that observed in numerical and experimental results concerning the continuous phase transitions in spin-1 BECs~\cite{Damski2007, Anquez2016}. 
As illustrated in Fig.~\ref{fig:Q_a-scaling}(b), the temporal scaling extends to other quantities, recovering the same $Q_a$ (see ``Appendix'').
The observed scaling is consistent with the Kibble-Zurek scaling presented above for our system. Taking $\nu = 1/2$ and $z=2$,  and using Eq.~\eqref{eq: KZ-freezing-scaling}, we obtain $\freeze{t} \sim \tau_Q^{1/2}$. 
Combining with the relation $Q_a=|\freeze{t}/\tau_Q|$, we also recover the $Q_a \sim \tau_Q^{-1/2}$ scaling seen in our simulations. 

To reinforce our conclusions, we also recover the scaling laws by directly linearising the GP equations around the critical point. In this case, the temporal dependence of the quadratic Zeeman terms is treated directly. 
Following Ref.~\cite{Damski2007}, we begin with a wave function close to the BA ground state, $\Psi^T = (\psi_1 + \delta\psi_1(t), \psi_0 + \delta\psi_0(t), \psi_{-1} + \delta\psi_{-1}(t))\exp(-i\mu t)$ where $\psi_{\pm 1}, \psi_0$ are defined in Eq.~\eqref{eq: BA-GS} with $Q=Q_0$.
Here, $\mu=c_0+c_1(2 - Q_0)/2$ is the chemical potential, $0 \leq Q_0 \leq 2$ is a constant, and $|\delta\psi_m(t)| \ll 1$.
The noise terms have to satisfy $\int \sum_m \delta\psi_m + \delta\psi_m^* \,  \dd z = 0$ to ensure the proper normalisation of the wave function and $\int (\delta\psi_1 + \delta\psi_1^* + \delta\psi_{-1} + \delta\psi_{-1}^*) \dd z = 0$ to enforce conservation of magnetisation.

Linearizing the spin-1 GPEs about the state corresponding to $Q=Q_0$ (see ``Appendix''),  we obtain
\begin{equation}
    \!\!\! i \hbar \dv{t}G_y = \left[-\frac{\hbar^2}{2M}\dv[2]{z} - c_1  n_0 \left(Q - \frac{Q_0}{2}\right)\right]G_y - \frac{c_1  n_0 Q}{2}G_y^*,
\end{equation}
where $G_y=\delta\psi_1-\delta\psi_{-1}$.
Next, we transform to momentum space and split $G_y$ into real and imaginary parts, 
where $a_y=\int \mathrm{Re}(G_y)e^{-ikz} \mathrm{d}z$ and $b_y=\int \mathrm{Im}(G_y)e^{-ikz} \mathrm{d}z$.
We then solve for $Q=-t/\tau_Q$, by deriving the equation for $\mathrm{d}^2a_y/\mathrm{d}t^2$ across a DQCP. Rescaling time as $t\rightarrow t\lambda$ with $\lambda = \sqrt{\tau_s\tau_Q}$, we arrive at
\begin{equation}
    \dv[2]{a_y}{t} = \frac{-1}{\left( 2\kappa^2 - t\right)}\dv{a_y}{t}-\frac{1}{4} \left( \kappa^4 - 2\kappa^2 t + \frac{3 t^2}{4} \right) a_y, \label{eqn_scaled_lin}
\end{equation}
where $\kappa^2 = \xi_s^2 k^2 \sqrt{\tau_Q/\tau_s}$. This scaling ensures that the last term is independent of $\tau_Q$. The remaining dependence on $\tau_Q$ is eliminated if we require that $\kappa$ is constant, which implies $k \sim \tau_Q^{-1/4}$. 
Only under these conditions can we expect scaling solutions, and they are also consistent with the scaling of the correlation length and the dynamical exponents derived earlier based on the KZM.

\subsection{Scaling of the number of domains}

\begin{figure}
    \centering
    \includegraphics[width=\columnwidth]{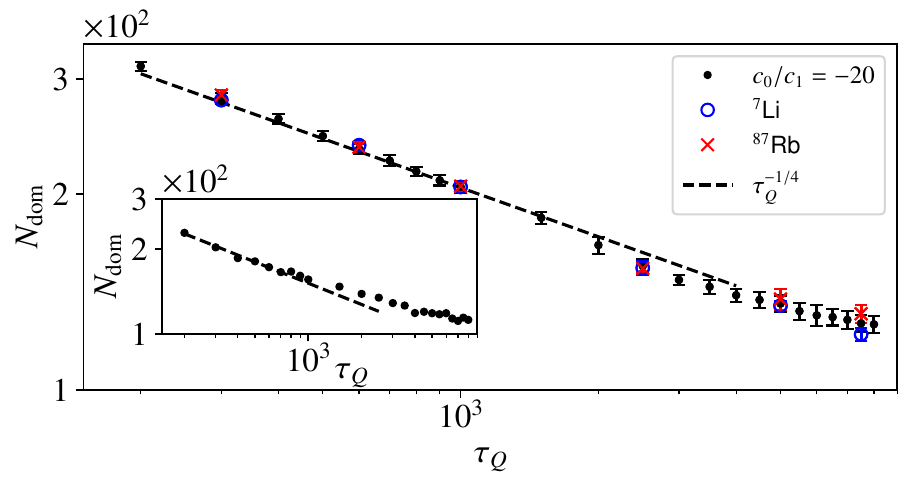}
    \caption{\textbf{Kibble-Zurek scaling of the number of domains.} The number of ferromagnetic domains (black dots) as a function of the quench time, $\tau_Q$.
    Overlaid is the power-law scaling $\tau_Q^{-1/4}$ (dashed line). The results are verified for both $^7$Li (blue circles) and $^{87}$Rb (red crosses) interaction parameters.
    Each point is averaged over 50 runs, and the error bars give $\pm 1$ standard deviation.
    Inset: number of ferromagnetic domains in a quench that spans two phase transitions for the same quench times as the
    main figure.}
    \label{fig: BA-FM-domains}
\end{figure}

The KZM not only predicts the growth of the correlation length immediately following the phase transition, but also the number of phase-separated, FM domains, whose formation in the long-time dynamics involves discontinuous changes in the properties of the system. This is a measurable quantity and has been investigated in works applying the KZM to continuous transitions~\cite{Sabbatini2011, Swislocki2013}. 
Here the domain formation occurs following a sudden change in the occupation-number density of the spinor components (Fig.~\ref{fig: density-deviation}). It is therefore a central question whether the KZM correctly predicts the scaling for the number of domains forming at late times.
We numerically determine the total number of FM domains (Fig.~\ref{fig: component-densities}) at the end of the simulations for a broad range of $\tau_Q$, as shown in Fig.~\ref{fig: BA-FM-domains} .
For sufficiently fast
quenches ($\tau_Q < 1000$),  a clear power-law scaling $N_{\text{dom}}\sim\tau_Q^{-1/4}$ emerges, which agrees 
well with Eq.~\eqref{eq: n_d-scaling} as well as the scaling obtained with Eq.~\eqref{eqn_scaled_lin}. 
As with $\freeze{t}$, the scaling for the DQCP is again different from that in analogous transitions across a continuous critical point~\cite{Sabbatini2011}.

Unlike $Q_a$, the number of domains shows a clear deviation from the predicted KZ scaling for slow quenches ($\tau_Q > 1000$).
Similar differences in the scaling of observables measured at much 
later times from the critical point have also been observed in \cite{Su2013,Swislocki2013}. In general, the scaling changes from power law to exponential decline in the limit when $N_{\text{dom}} \sim 1$, as predicted, e.g., by the Landau-Zener model~\cite{Zurek2005}. Here we find that a deviation occurs already for intermediate values of $\tau_Q$ and $N_{\text{dom}} \sim 100$. Based on similar behaviour in spin chains~\cite{Pellegrini2008,Divakaran2008} and exact results for the Ising model~\cite{Dziarmaga2010}, this could be attributed to effects of finite-size scaling or the presence of another, undetermined, length scale that dominates for intermediate $\tau_Q$.

\begin{figure}
    \centering
    \includegraphics[width=\columnwidth]{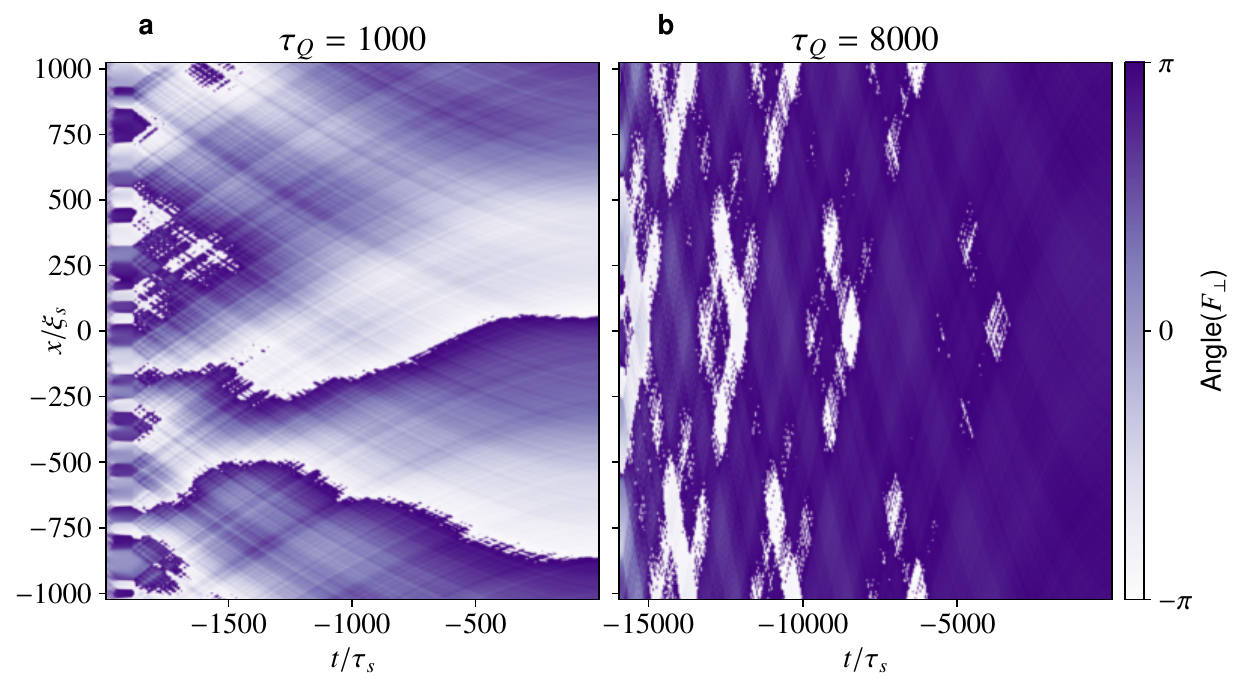}
    \caption{\textbf{Phase jumps induced by quenching across two quantum phase transitions.} Angle of transverse magnetisation (colour scale) following crossing of the polar-to-BA phase transition for (a) a fast, and (b) a slow quench.  In the former case, two discernible phase jumps appear and persist after the transition (but are absent in the latter).}
    \label{fig:magnetization-angle}
\end{figure}

Finally, we confirm the robustness of the scaling by considering a quench that crosses two phase transitions, starting from the polar state with $Q=2.5$, quenching through the second-order transition into the BA phase and then continuing across the BA-FM DQCP. This scenario is of particular importance, since it may be experimentally simpler to realise. We find that the crossing of the polar-to-BA phase transition can excite modes associated with the transverse magnetisation ($F_{\perp}$), as found for sudden quenches~\cite{Schmied2019,Siovitz2023}. These excitations can be seen in Fig.~\ref{fig:magnetization-angle} which depicts the angle (direction) of $F_{\perp}$ following the phase transition to the BA phase for two quench rates. Note that these excitations can give rise to phase jumps in the angle of $F_{\perp}$ that can persist beyond the transition region.
To determine the impact on the KZ scaling behaviour, we show the number of domains also for the two-QPT scenario in the inset of Fig.~\ref{fig: BA-FM-domains}. The results remain qualitatively similar with the same $\tau_Q^{-1/4}$ scaling and deviation for slow quenches. We have also verified that after the polar-to-BA transition, the system emerges from the impulse regime and tracks the BA phase with the aforementioned excitations present before entering a second impulse regime associated with the DQCP. Therefore, our results correspond to two well-separated impulse regimes and demonstrate the reproducibility of our scaling law for a range of initial conditions.

\section{Conclusion}

In conclusion, 
we have shown that the KZM can be generalised to this discontinuous phase transition, leading to scaling laws that differ from those observed for phase transitions across continuous quantum critical points for the same spin-1 BEC model. We find excellent agreement with numerical simulation for both the short-time growth of the unstable excitations and the subsequent number of domains formed on longer time scales. Our results hold for experimentally accessible parameter regimes allowing these extensions of the KZM to be realized in current experiments on spinor BECs, which therefore emerge as prime candidates for testbed systems for investigating critical scaling in first-order quantum phase transitions, including as laboratory emulators for understanding false-vacuum decay~\cite{Billam2022,Song2022,Zenesini2023,Lagnese2023}.

\begin{acknowledgments}
The numerical simulations were carried out on the High Performance Computing Cluster supported by the Research and Specialist Computing Support service at the University of East Anglia.  MOB acknowledges support from Engineering and Physical Sciences Research Council under Grant No.\ EP/V03832X/1.
\end{acknowledgments}

\appendix*

\section{}

\subsection{Numerical Simulation}
We measure length and time in units of the spin healing length $\xi_s=\hbar/\sqrt{2M|c_1|n_0}$ and the spin time $\tau_s = \hbar/2|c_1|n_0$, respectively.
Our simulations are performed on a 1D grid of $N_x = 16384$ points with a spacing of $\Delta_x=0.125\xi_s$,
considering a ring-shaped geometry by assuming periodic boundary conditions and $V(z) = 0$.
We start from Eq.~\eqref{eq: BA-GS}, adding small noise terms, $\delta\psi_m$, to each component, where $|\delta\psi_m| \ll 1$.
The real and imaginary parts of $\delta\psi_m$ are drawn from the probability distribution $p(z) = \exp(-z^2/2\sigma^2)/(\sqrt{2\pi}\sigma)$, 
with $\sigma=10^{-4}$ to remain close to the BA ground state.
We vary the quadratic Zeeman shift as $Q(t) = -t/\tau_Q$ for a range of quench times $\tau_Q$, starting at $Q=1$ and ending the simulation at $Q=-2.5$.

\subsection{Bogoliubov modes for the broken-axisymmetry phase}
The broken-axisymmetry (BA) phase of a spin-1 BEC exhibits three Bogoliubov modes~\cite{Uchino2010}.
Here we rederive each mode explicitly from the relevant Bogoliubov
transformations and explain why $E_{\vb{k}, f_z}$
is the relevant mode for the BA-to-ferromagnetic (FM) transition.

The broken-axisymmetry phase can be parameterised as
\begin{equation}
    \zeta^\text{BA} = \left(\frac{\sin\theta}{\sqrt{2}}, \cos\theta, \frac{\sin\theta}{\sqrt{2}}\right),
\end{equation}
where $\sin\theta = \sqrt{1/2+q/(4nc_1)}$.
The fluctuation operators for this state are then defined as~\cite{Uchino2010}:
\begin{align}
  \hat{a}_{\vb{k}, d} &= \frac{\sin\theta}{\sqrt{2}}(\hat{a}_{\vb{k}, 1} + \hat{a}_{\vb{k}, -1}) + \cos\theta \hat{a}_{\vb{k}, 0}, \\
  \hat{a}_{\vb{k}, f_z} &= \frac{1}{\sqrt{2}}(\hat{a}_{\vb{k}, 1} - \hat{a}_{\vb{k}, -1}), \\
  \hat{a}_{\vb{k}, \theta} &= \frac{\cos\theta}{\sqrt{2}}(\hat{a}_{\vb{k}, 1} + \hat{a}_{\vb{k}, -1}) - \sin\theta \hat{a}_{\vb{k}, 0},
\end{align}
where on the right-hand side $\hat{a}_{\vb{k}, m}$ is the annihilation operator for a
spin-$1$ boson in magnetic level $m$ (for $m=-1,0,+1$), determined by expanding the wave-function field operator as
\begin{equation}
    \hat{\psi}_m(\vb{x}) = \frac{1}{\sqrt{V}}\sum_{\vb{k}}
    \hat{a}_{\vb{k}, m}e^{i\vb{k}\cdot \vb{x}},
\end{equation}
where $V$ is the volume of the system.

The sub-Hamiltonian for the spin fluctuation mode $\hat{a}_{\vb{k}, f_z}$ can be diagonalized using the transformation
\begin{equation}
    \hat{b}_{\vb{k}, f_z} = \sqrt{\frac{\epsilon_{\vb{k}} + q/2 + E_{\vb{k}, f_z}}{2E_{\vb{k}, f_z}}}\hat{a}_{\vb{k}, f_z}
    + \sqrt{\frac{\epsilon_{\vb{k}} + q/2 - E_{\vb{k}, f_z}}{2E_{\vb{k}, f_z}}}\hat{a}_{-\vb{k}, f_z}^\dagger,
\end{equation}
where $\epsilon_{\vb{k}} = \hbar^2|\vb{k}|^2/2M$ is the kinetic energy and the Bogoliubov
spectrum is given by
\begin{equation}\label{eq: e_k-fz}
    E_{\vb{k}, {f_z}} = \sqrt{\epsilon_{\vb{k}}(\epsilon_{\vb{k}} + q)}.
\end{equation}
The sub-Hamiltonians for the density fluctuation mode $\hat{a}_{\vb{k}, d}$ and the $\theta$ mode
$\hat{a}_{\vb{k}, \theta}$ can be similarly diagonalized using operators $\hat{b}_{\vb{k}, +}$ and
$\hat{b}_{\vb{k}, +}$, 
which yields the remaining two Bogoliubov modes~\cite{Uchino2010}:
\begin{widetext}
  \begin{equation}\label{eq: e_k-pm}
    E_{\vb{k}, \pm}= \sqrt{\epsilon_{\vb{k}}^2 + n(c_0-c_1)\epsilon_{\vb{k}}
      + 2{(nc_1)}^2(1 - \tilde{q}^2) \pm E_1(\vb{k})},
  \end{equation}
  where \(\tilde{q} = -q/2c_1n\) and
  \begin{equation}
    E_1(\vb{k}) = \sqrt{{\left[n{(c_0 + 3c_1)}\epsilon_{\vb{k}}
          + 2{(c_1n)}^2(1-\tilde{q}^2)\right]}^2
      - 4c_1(c_0+2c_1){(n\tilde{q}\epsilon_{\vb{k}})}^2}.
  \end{equation}
The final, diagonalized Hamiltonian then reads
\begin{equation}
    \hat{H}^\text{BA} =\, E_0^\text{BA}
    + \sum_{\vb{k} \neq 0}[E_{\vb{k}, f_z}\hat{b}^\dagger_{\vb{k}, f_z}\hat{b}_{\vb{k}, f_z} 
    + E_{\vb{k}, -}\hat{b}^\dagger_{\vb{k}, -}\hat{b}_{\vb{k}, -}
    + E_{\vb{k}, +}\hat{b}^\dagger_{\vb{k}, +}\hat{b}_{\vb{k}, +}],
\end{equation}
where $E_0^\text{BA}$ is the ground state energy for the
BA phase, which is explicitly derived in Ref.~\cite{Uchino2010}.
\end{widetext}

We now consider our 1D system.
In the long-wavelength limit, \(k \rightarrow 0\), the only non-zero
(gapped) mode is \(E_{k, +} = \sqrt{4{(c_1n)}^2(1-\tilde{q}^2)}\) which has the form
\(E_{k, +} \sim \sqrt{{q_c^\prime}^2 - q^2}\) with \(q_c^\prime=2c_1n\).
The relevant mode of the BA to FM transition is found from the imaginary parts of the Bogoliubov energies (Fig~\ref{fig: bogoliubov-energies}). For \(|Q| > 2\), where \(Q \equiv q / (|c_1|n)\), \(E_{k, +}\) has a positive imaginary part, indicating
instability.
The critical point \(Q = 2\) (\(q=q_c^\prime\)) corresponds to the second-order transition between the polar and BA phases, for which \(E_{k, +}\) therefore is the corresponding Bogoliubov energy.

However,  for \(Q < 0\) the imaginary part of \(E_{k,f_z}\) mode becomes non-zero and positive,
and thus unstable.
This corresponds precisely to  the transition from the BA to the FM
phase at \(Q=0\) that we are interested in here, and therefore \(E_{k, f_z}\) is the relevant mode to study.
Note that the $E_{k, f_z}$ mode does not give rise to instability at $k = 0$.
Therefore, studies focusing on this mode at $k=0$ only do not capture the phase transition that occurs at $Q=0$~\cite{Matuszewski2009, Qiu2020, Mirkhalaf2021}.
In contrast, the $k=0$ mode corresponds to the most unstable mode for $E_{k, +}$,
and thus it suffices to choose this Bogoliubov energy to capture the phase transition that
occurs at $Q=2$.
In practice, the $Q=-2$ transition is not realized since the instability of \(E_{k,f_z}\) at any $k \ne 0$ will typically arise at $Q=0$ when $Q$ is quenched from positive to negative values. 

\subsection{Extracting the freezing time}

In order to extract the freezing time $\tilde{t}$ of the system, an appropriate quantity
must be chosen.
Since the transition to the FM phase causes the formation of phase-separated
domains, a natural choice is to measure fluctuations in the difference of
the populations of \(\psi_{\pm 1}\)~\cite{Uchino2010}.
To do this, we first construct the Fourier transforms of the \(\psi_{\pm 1}\)
components as \(\hat{a}_{\pm 1}(k) = \int \psi_{\pm 1}
e^{-ikz} \, d\vb{z}\).
After passing through the critical point into the FM phase, the difference \(\hat{a}_{k, f_z}(k) = \hat{a}_1(k)
- \hat{a}_{-1}(k) / \sqrt{2}\)
generates measurable fluctuations as FM domains with opposite spin start to form [see inset of Fig.~\ref{fig:Q_a-scaling}(a)], while before the transition it remains zero (due to the
absence of domains).
To measure the freezing time, we extract the time at which
\(|\hat{a}_{k, f_z}(k)|\) exceeds some appropriately chosen value.
In our numerical simulations, we take this value to be 1\% of the maximum value
of \(|\hat{a}_{k, f_z}(k)|\) over the entire simulation.

An alternative choice is the population of the \(\psi_0\) component.
Here, instead of measuring the growth of a quantity, we now extract the
freezing time as the time required for the \(\psi_0\) component to deviate
from its analytically calculated value in the (metastable) BA phase \(\psi_0 = \sqrt{2+Q} / 2\) (Fig.~\ref{fig: density-deviation}).
In particular, we choose the freezing time to be the time at which the
deviation reaches 1\% of the analytically predicted value, yielding Fig.~\ref{fig:Q_a-scaling}(b).
We see that, despite using an entirely different quantity to measure the
freezing time, the resulting scaling of \(Q_a\) is the same as when calculated from $|\hat{a}_{k,f_z}(k)|$.

\subsection{Deriving scaling near the critical point}
The spin-1 Gross-Pitaevskii equations (GPEs) are given as~\cite{Kawaguchi2012}:
\begin{equation}\label{eq: GPEs}
    i\hbar\pdv{\Psi}{t} = \left[ -\frac{\hbar^2\nabla^2}{2M}
    -p\hat{F_z} + q\hat{F}_z^2 +c_0n +c_1n\langle \hat{\vb{F}}\rangle \cdot \hat{\vb{F}}\right]\Psi.
\end{equation}
Recall that we start from a BA phase of the form $\Psi^T = (\psi_1
+ \delta\psi_1(t), \psi_0 + \delta\psi_0(t), \psi_{-1}
+ \delta\psi_{-1}(t))\exp(-i\mu t)$.
Substituting this expression into the GPEs and keeping leading order
terms in $\delta\psi_m$ yields the following equations for $\delta\psi_{\pm 1}$ ($p=0$)
\begin{widetext}
  \begin{equation}\label{eq: dpsi1}
    \begin{split}
      i\hbar\pdv{\delta\psi_1}{t} = &\left[-\frac{\hbar^2}{2M}\dv[2]{}{z}+q-\mu
        + \frac{n_0(10c_0+6c_1-(c_0-c_2)Q)}{8}\right]\delta\psi_1
      + \frac{n_0\sqrt{2(4-Q^2)}}{8}\left[
        (c_0 + 3c_1)\delta\psi_0
        + (c_0+c_1)\delta\psi_0^*
      \right] \\
      &+ \frac{n_0(2-Q)}{8}\left[(c_0-c_1)\delta\psi_{-1}
        + (c_0+c_1)\delta\psi_1^*\right]
      + \frac{n_0}{8}\left[(2-Q)c_0 + (2+3Q)c_1\right]\delta\psi_{-1}^*,
    \end{split}
  \end{equation}
  \begin{equation}\label{eq: dpsim1}
    \begin{split}
      i\hbar\pdv{\delta\psi_{-1}}{t} = &\left[-\frac{\hbar^2}{2M}\dv[2]{}{z}+q-\mu
        + \frac{n_0(10c_0+6c_1-(c_0-c_2)Q)}{8}\right]\delta\psi_{-1}
      + \frac{n_0\sqrt{2(4-Q^2)}}{8}\left[
        (c_0 + 3c_1)\delta\psi_0
        + (c_0+c_1)\delta\psi_0^*
      \right] \\
      &+ \frac{n_0(2-Q)}{8}\left[(c_0-c_1)\delta\psi_{1}
        + (c_0+c_1)\delta\psi_{-1}^*\right]
      + \frac{n_0}{8}\left[(2-Q)c_0 + (2+3Q)c_1\right]\delta\psi_{1}^*.
    \end{split}
  \end{equation}
  Subtracting Eq.~\eqref{eq: dpsim1} from Eq.~\eqref{eq: dpsi1} results
  in the differential equation for $G_y = \delta\psi_1 - \delta\psi_{-1}$:
  \begin{equation}\label{eq: G_y-unsimplified}
    i\hbar\dv{G_y}{t}= \left[ -\frac{\hbar^2}{2M}\dv[2]{}{z}+q-\mu+n_0(c_0+c_1)\right]G_y
    -\frac{c_1n_0Q}{2}G_y^*.
  \end{equation}
  Additionally, to calculate the chemical potential, we take the
  $\psi_0$ component of Eq.~\eqref{eq: GPEs} keeping lead order
  terms and assuming a stationary state, which leads to $\mu = c_0n_0
  + c_1n_0(2-Q_0)/2$ where $Q_0$ is a constant.
  Substituting this expression and $q(t)=-c_1n_0Q(t)$ into Eq.~\eqref{eq: G_y-unsimplified} yields
  \begin{equation}\label{eq: G_y-simplified}
    i\hbar\dv{G_y}{t} = \left[ -\frac{\hbar^2}{2M}\dv[2]{}{z}
      -c_1n_0\left(Q-\frac{Q_0}{2}\right)\right]G_y
    - \frac{c_1n_0Q}{2}G_y^*.
  \end{equation}
  
  To progress, we split $G_y$ into real and imaginary parts and
  take the Fourier transform: $a_y=\int \mathrm{Re}(G_y)e^{-ikz} \mathrm{d}z$
  and $b_y=\int \mathrm{Im}(G_y)e^{-ikz} \mathrm{d}z$.
  Substituting into Eq.~\eqref{eq: G_y-simplified} yields the 
  matrix equation
  \begin{equation}\label{eq: a_y-matrix}
    \!\!\!\!\! \dv{t}\mqty[a_y \\ b_y] = \mqty(0 & \frac{\hbar k^2}{2M} - \frac{c_1  n_0}{2\hbar}(Q - Q_0) \\
    \frac{c_1  n_0}{2\hbar} \left( 3Q-Q_0 \right)-\frac{\hbar k^2}{2M} & 0)\mqty[a_y \\ b_y].
  \end{equation}
  To solve the above equation, we construct the equation for $\dv[2]{a_y}{t}$ and take $Q_0=0$, which yields
  \begin{equation}
    \dv[2]{a_y}{t} = \frac{c_1n_0}{2\hbar\tau_Q}b_y + \left(\frac{\hbar^2k^2}{2M} - \frac{c_1n_0Q}{2\hbar}\right)\dv{b_y}{t}.
  \end{equation}
  Expressions for $b_y$ and $db_y/dt$ are found from Eq.~\eqref{eq: a_y-matrix}.
  Substituting these in yields the following equation for $d^2a_y/dt^2$:
  \begin{equation}
    \dv[2]{a_y}{t} = \frac{1}{\tau_Q\left(\frac{\hbar^2k^2}{Mc_1n_0} - Q\right)}\dv{a_y}{t}
    - \left(\frac{\hbar^2k^4}{4M^2} - \frac{k^2c_1n_0Q}{M} + \frac{3c_1^2n_0^2Q^2}{4\hbar^2}\right)a_y.
  \end{equation}
  To simplify the above expression, we use the spin healing length $\xi_s = \hbar/\sqrt{2|c_1|n_0}$ and the spin time $\tau_s=\hbar/|c_1|n_0$:
  \begin{equation}
    \!\!\!\!\!\!  \dv[2]{a_y}{t} = \frac{-1}{\left(2\xi_s^2 k^2 \tau_Q- t \right)}\dv{a_y}{t}
    -\left( \frac{\xi_s^4 k^4}{4\tau_s^2} -\frac{\xi_s^2 k^2 t}{2\tau_s^2 \tau_Q} + \frac{3 t^2}{16\tau_s^2 \tau_Q^2} \right) a_y.
  \end{equation}
  Rescaling time as $t\rightarrow t\lambda$ with $\lambda = \sqrt{\tau_s\tau_Q}$, leads to the differential equation
  \begin{equation}
    \dv[2]{a_y}{t} = \frac{-1}{\left( 2\kappa^2 - t\right)}\dv{a_y}{t}-\frac{1}{4} \left( \kappa^4 - 2\kappa^2 t + \frac{3 t^2}{4} \right) a_y, \label{eqn_scaled_lin_methods}
  \end{equation}
  where $\kappa^2 = \xi_s^2 k^2 \sqrt{\tau_Q/\tau_s}$.
\end{widetext}

\noindent{\textbf{Data availability}}

\noindent The data for this study are available at \href{https://doi.org/10.5281/zenodo.14996614}{https://doi.org/10.5281/zenodo.14996614}

\smallskip
\noindent{\textbf{Code availability}} 

\noindent The code is available from the authors upon reasonable request.

\smallskip
\noindent{\textbf{Author contributions}} 

\noindent M.O.B suggested and developed the project with contributions from H.S. M.T.W.\ wrote the code and performed the numerical simulations. M.T.W.\ and H.S.\ developed the analytical derivations with input from M.O.B. All authors contributed to discussion and analysis of numerical data, analytical theory, and interpretation. The manuscript was written jointly.

\smallskip
\noindent{\textbf{Competing interests}}

\noindent The authors declare no competing financial or non-financial interests.

\bibliography{references_abbr}

\end{document}